\documentclass[pra,twocolumn,showpacs,superscriptaddress,amsfonts,amsmath,
floatfix]{revtex4}
\usepackage{graphicx}
\usepackage{psfrag}
\newcommand{\Journal}[4]{#1 {\bf #2}, #3 (#4)}
\newcommand{\PR}{Phys. Rev.}
\newcommand{\PRL}{Phys. Rev. Lett.}

\newcommand{\JMP}{J. Math. Phys.}

\newcommand{\Science}{Science}

\newcommand{\PLA}{Phys. Lett. A}
\begin{document}
\title {Pairing of a harmonically trapped fermionic Tonks-Girardeau gas}
\author{A. Minguzzi}
\email{anna.minguzzi@grenoble.cnrs.fr}
\affiliation{Laboratoire de Physique et Mod\'elisation des Mileux Condens\'es, 
C.N.R.S., B.P. 166, 38042 Grenoble, France}
\affiliation{Laboratoire de Physique Th\'{e}orique et Mod\`{e}les
Statistiques, Universit\'{e} Paris-Sud, B\^{a}t. 100, F-91405 Orsay, France}
\author{M. D. Girardeau}
\email{girardeau@optics.arizona.edu}
\affiliation{College of Optical
Sciences, University of Arizona,
Tucson, AZ 85721, USA}
\date{\today}
\begin{abstract}
The fermionic
Tonks-Girardeau (FTG) gas is a one-dimensional spin-polarized Fermi
gas  with infinitely strong attractive zero-range   
odd-wave interactions,  arising from a confinement-induced resonance reachable 
via a three-dimensional  $p$-wave Feshbach resonance.
We investigate the off-diagonal long-range order (ODLRO) of the FTG gas
subjected to a longitudinal harmonic confinement by analyzing the 
 two-particle reduced density matrix
for which we derive  a closed-form 
expression. Using a variational 
approach and numerical diagonalization we find that the largest eigenvalue of the two-body
density matrix is of order $N/2$, where $N$
is the total particle number, and hence a partial ODLRO is present for
a FTG gas in the trap.
\end{abstract}
\maketitle

\section{Introduction}
Low-dimensional systems display unusual and striking features as
compared to their three-dimensional (3D) counterparts, among which are
the enhanced effects of the interactions and the presence of large
fluctuations which are responsible for the failure of mean-field
approaches. In addition, especially in the one-dimensional (1D) case,
it is possible to find exact solutions which greatly help the progress
in our understanding of complex, strongly-interacting many-body
systems.

Low-dimensional atomic quantum gases are one of the frontiers of the
current theoretical and experimental investigations. Bosonic and
fermionic atomic gases constrained to a quasi-1D geometry
have already been realized experimentally by trapping atoms in
two-dimensional optical lattices \cite{Gre01,Gun05}. In the case of
bosons the strongly-repulsive (known as Tonks-Girardeau) 
regime has been achieved
\cite{Par04,Kin04}. This regime has been fully understood thanks to the
knowledge of the exact many-body wavefunction through the mapping of
the strongly repulsive, impenetrable bosons onto an ideal gas of
fermions subjected to the same external potential \cite{Gir60}. Quite
remarkably, the mapping does not hold only for the ground state, but
may be extended to treat time-dependent phenomena and systems at
finite temperature \cite{GirWri00a,GirWri00b,Kas02}.

In this paper we focus on the model of spin-polarized fermions
interacting via strongly attractive odd-wave interactions, which are
the 1D analogue of $p$-wave interactions in 3D. This regime might be
experimentally reachable by exploiting the so-called
confinement-induced resonances (CIR), which permit to tune the
1D coupling constant via a 3D Feshbach resonance \cite{GraBlu04}. In the
limit of infinitely strong attractions, known as the fermionic
Tonks-Girardeau (FTG) regime, the many-body wavefunction is known
exactly through an inverse Fermi-Bose mapping which allows to express
the fermionic wavefunction in terms of the one of an ideal Bose
gas \cite{GraBlu04,GirOls03,GirNguOls04}. At the resonance point we can thus
quantitatively address the 
question of what the structure of the ground state is, and in
particular whether the fermions are paired. This contributes to the
understanding of the intermediate, strongly-interacting region in
 the 1D equivalent of the BCS-BEC
crossover for $p$-wave fermions.

In a previous work \cite{GirMin05} we have studied the issue of pairing
for a homogeneous FTG gas in the thermodynamic limit. We have  found that there
is off-diagonal long-range order (ODLRO) in the reduced two-body density
matrix, indicating a paired state. We consider here the
experimentally-relevant case of a gas with a finite number of particles and  
subjected to a longitudinal
harmonic confinement. We derive an an exact analytic expression for
the two-body density matrix of the trapped gas and both by variational
estimates and by numerical diagonalization we show that also in the
presence of the trap ODLRO persists, yielding a complex, partially
paired quantum state.

\section{Fermionic Tonks-Girardeau gas in a longitudinal harmonic trap} 
We consider an atomic spin-polarized Fermi gas confined by a tight
atom waveguide of which it occupies the transverse ground state. We
shall henceforth assume that the level spacing in the transverse
direction is much larger than all the relevant energy scales in the
problem (such as chemical potential, temperature, etc.), so that the
gas is effectively one-dimensional. We also consider the
case of a gas subjected to a much weaker
longitudinal harmonic confinement of frequency $\omega$.  
The 1D Hamiltonian is
\begin{equation}\label{eq1}
\hat{H}=\sum_{j=1}^{N}
\left[-\frac{\hbar{^2}}{2m}\frac{\partial^2}{\partial x_{j}^{2}}\right]
+\sum_{1\le j<\ell\le N}v_\text{int}^{\text{F}}(x_{j}-x_{\ell})+\sum_{j}\frac{1}{2}m\omega^{2}x_{j}^2
\end{equation}
where $v_\text{int}^{\text{F}}$ is a short-range attractive two-body
interaction 
which is specified here below. Since the spatial
wave function is antisymmetric due to the spin polarization, there is no 
zero-range $s$-wave (delta function) interaction, but it has been shown
\cite{GraBlu04,GirOls03,GirNguOls04} that a strong, attractive, and 
short-range odd-wave interaction (a 1D analog of 3D $p$-wave interactions) 
occurs in the neighborhood of the CIR. Such an interaction can be
expressed through a contact condition for the relative
wavefunction $\psi_{F}(x)$ of each pair of fermions as
\cite{CheShi98}
\begin{equation}
\label{eq:contact}
\left. \frac{1}{\psi_{F}(x)}\frac{d
  \psi_{F}(x)}{dx}\right|_{x=0^+}=-\frac{2\hbar^2}{m g_{1D}^F},
\end{equation}
where $g_{1D}^F$ is the 1D fermionic coupling constant, which can be
expressed in terms of the 3D $p$-wave scattering volume \cite{GraBlu04}.
The condition (\ref{eq:contact}) above, together with the antisymmetry condition
$\psi_{F}(x<0)=-\psi_{F}(x>0)$ on the relative 
wavefunction  leads to a solution for
$\psi_{F}(x)$ which is discontinuous at the contact point $x=0$ but has a
continuous first derivative. 

 The FTG gas corresponds to the negative side of the CIR, that is, the
  case where $g_{1D}^F\rightarrow -\infty$; in the FTG limit  the first
  derivative of the relative  wavefunction $\psi_{F}(x)$  equals to zero
  at $x=0$.
For the sake of illustration, let us consider first two particles under harmonic confinement. In the presence of such a confinement  the Schroedinger equation can be solved exactly  for any value of the coupling strength $g_{1D}^F$ \cite{Busch98}. The relative wavefunction 
$\psi_{F}(x)$ in the FTG limit 
  takes the value
  $\psi_{F}(x>0)=(2\pi)^{-1/4}x_{osc}^{-1/2}e^{-Q^{2}/4}$ and $\psi_{F}(x<0)=-(2\pi)^{-1/4}x_{osc}^{-1/2}e^{-Q^{2}/4}$,  where we have set 
  $Q=x/x_{osc}$   and $x_{osc}=\sqrt{\hbar/m\omega}$.  

The solution for $N=2$ can  be extended to arbitrary particle numbers $N$, so that the exact fermionic TG gas ground-state wavefunction is \cite{GirOls03,GraBlu04}
\begin{equation}
\label{eq:psiF}
\Psi_{F}(x_{1},\cdots,x_{N})=
A(x_{1},\cdots,x_{N})\prod_{j=1}^{N}\phi_{0}(x_{j})
\end{equation} 
 with
where $A(x_{1},\cdots,x_{N})=\prod_{1\le j<\ell\le
N}\text{sgn}(x_{\ell}-x_{j})$ is the ``unit antisymmetric function''
employed in the original discovery of fermionization \cite{Gir60} and
$\phi_0(x)=\pi^{-1/4}x_{osc}^{-1/2}e^{-Q^{2}/2}$ is  the
  ground-state 
  orbital of the longitudinal harmonic confinement.
 Hence, the FTG gas
  is mapped through the function $A$ to the ground state of an ideal
  Bose gas under harmonic confinement, of which it shares all the
  properties that do not depend on the sign of the
  many-body wavefunction, such as the density profile and the spectrum
  of collective excitations.

\section{Off-diagonal long-range order for a trapped FTG gas}

In order to explore the pairing properties of the FTG gas we study the
two-body reduced density matrix, defined as 
 \begin{eqnarray}
\label{defrho2}
\rho_{2}(x_{1},x_{2};x_{1}',x_{2}')&
=& N(N-1)\int \Psi_{F}(x_{1},x_{2}\cdots,x_{N})\nonumber \\ &\times &
\Psi_{F}^{*}(x_{1}',x_{2}',x_{3},\cdots,x_{N}) dx_{3} \cdots 
dx_N. \nonumber \\
\end{eqnarray}
As it was discussed by Yang \cite{Yan62}, the criterion for
``superconductive'' off-diagonal long-range order (ODLRO) in a
trapped system is that  the largest  eigenvalue $\lambda_1$ of the two-body density matrix  is  of
the order of the number of particles $N$, i.e. $\lambda_1=\alpha N$, with $0<\alpha \le 1$ being the pair-condensate fraction.

Using the exact form (\ref{eq:psiF}) of the many-body wavefunction, the integration over $N-2$ variables in Eq.~(\ref{defrho2}) can be explicitly performed and  we
obtain an analytic expression for the two-body density matrix, which
reads
\begin{eqnarray} 
\label{eq:rho2} 
& &\rho_{2}(x_{1},x_{2};x_{1}',x_{2}')
= N(N-1)\text{sgn}(x_{1}-x_{2})\phi_{0}(x_{1})\phi_{0}(x_{2})\nonumber\\ 
&\times&\text{sgn}(x_{1}'-x_{2}')
\phi_{0}(x_{1}')\phi_{0}(x_{2}') 
[G(x_{1},x_{2};x_{1}',x_{2}')]^{N-2},
\end{eqnarray} 
where 
$G(x_{1},x_{2};x_{1}',x_{2}')
=1+
\text{erf}(y_1)-\text{erf}(y_2)+\text{erf}(y_3)-\text{erf}(y_4)$, and
$y_1\le y_2\le y_3\le y_4$ are 
$(Q_{1},Q_{2},Q_{1}',Q_{2}')$ in ascending order, and   $Q_i=x_i/x_{osc}$. 
  
\begin{figure}
  \centering
 \psfrag{N}{$N$} 
\psfrag{l1}{$\lambda_1$}
\includegraphics[width=7.5cm,angle=0]{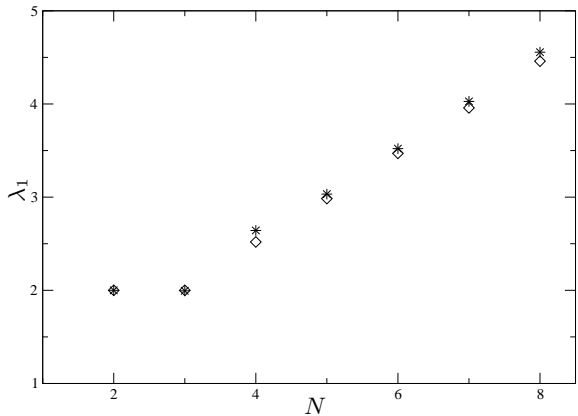} 
  \caption{Largest eigenvalue $\lambda_1$ of the
  two-body density matrix  for a harmonically trapped fermionic TG gas at
  various particle numbers $N$ from numerical diagonalization (stars)
  and the variational  lower-bound estimate (diamonds).}
  \label{fig1}
\end{figure}

\begin{figure}
\centering
\psfrag{fig1x}{$x_{1}/x_{osc}$}
\psfrag{fig1y}{$u_{1}(x_{1},0+)/u_{1}(0+,0-)$}
\includegraphics[width=7.5cm,angle=0]{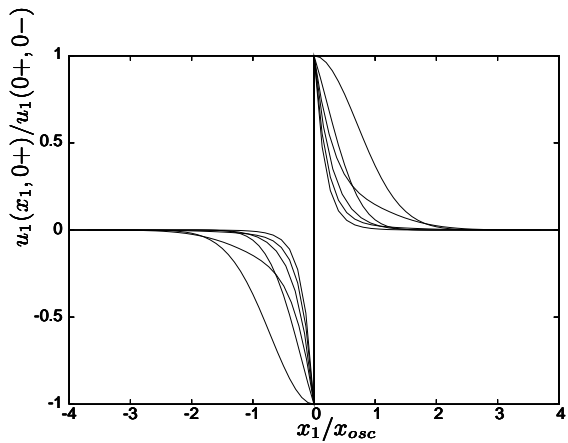} 
  \caption{Eigenfunctions of the two-body density matrix of a harmonically 
trapped fermionic TG gas belonging to the largest eigenvalue $\lambda_1$, 
for $N=2$ to $7$ from top to bottom (when $x_1>0$).}
  \label{fig2}
\end{figure}

The eigenvalues $\lambda_j$ and eigenfunctions $u_j$ 
of the two-body density matrix  are the solutions of the integral equation 
 \begin{equation}
\label{eq:eig}
\int dx_1'\,dx_2'\rho_2(x_{1},x_{2};x_{1}',x_{2}')u_j(x_{1}',x_{2}')
= \lambda_j u_j(x_{1},x_{2}). 
\end{equation}
If $N=2$ 
then $G(x_{1},x_{2};x_{1}',x_{2}')=1$, the two-body density matrix separates trivially and has only one nonzero eigenvalue
$\lambda_1=N=2$, with eigenfunction 
$u_1(x_1,x_2)=\text{sgn}(x_{1}-x_{2})\phi_{0}(x_{1})\phi_{0}(x_{2})$.
For $N=3$ we have found by numerical integration that the ansatz $u_1(x_1,x_2)
=\mathcal{C}\ \text{sgn}(x_{1}-x_{2})\phi_{0}(x_{1})\phi_{0}(x_{2})
[1-|\text{erf}(Q_1)-\text{erf}(Q_2)|]$ satisfies the above eigenvalue equation 
to within $0.1\%$ with eigenvalue $\lambda_1=2$ (the same as for $N=2$), and we conjecture that these expressions are exact.   
For $N\ge 4$ we have numerically solved the
eigenvalue equation (\ref{eq:eig}) by diagonalization. The results for the
largest eigenvalue are displayed in Fig.~\ref{fig1}
for $N=4$ to $8$.  
 The obtained values agree with the analytical expression 
 $\lambda_{1}=(N+1)/2$ to within the computational
accuracy of a few percent \cite{Note4}, indicating that in the trap a partial ODLRO is present.
%
The eigenfunctions corresponding to the largest eigenvalues obtained
by numerical diagonalization are  
shown in Fig. \ref{fig2} along with the analytical eigenfunctions 
for $N=2$ and $3$. 

\begin{figure}
  \centering
\psfrag{fig2x}{$x_1/x_{osc}$}
\psfrag{fig2y}{$u(x_{1},0+)/u_{1}(0+,0-)$}
\includegraphics[width=7.5cm,angle=0]{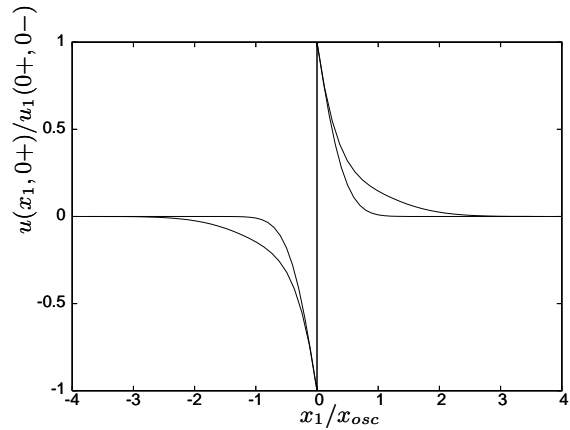} 
  \caption{Comparison of maximal eigenfunction $u=u_1$ of the two-body density 
matrix (upper curve when $x_{1}>0$) and variational ansatz $u=u_{trial}$ 
(lower curve when $x_{1}>0$) for the case $N=4$. In the case $N=3$
(not plotted) the two curves (on the same scale) would be indistinguishable.
}
  \label{fig3}
\end{figure}

We have also calculated lower bounds to $\lambda_1$ for $N=4$ to $8$ 
from the inequality
\begin{equation} 
\lambda_{1}\ge\frac{\int u_{trial}({\mathbf X})\rho_2({\mathbf X},{\mathbf X'})
u_{trial}({\mathbf X'})d{\mathbf X} d{\mathbf X}'}{\int u_{trial}^2({\mathbf X})d{\mathbf X}},
\end{equation}
where ${\mathbf X}$ is a shorthand notation  for $(x_1,x_2)$.
In particular, we have chosen the ansatz 
$u_{trial}(x_1,x_2)
=\mathcal{C}\ \text{sgn}(x_{1}-x_{2})\phi_{0}(x_{1})\phi_{0}(x_{2})
[1-|\text{erf}(Q_1)-\text{erf}(Q_2)|]^{N-2}$. 
As in is illustrated in Fig.~\ref{fig1}, the results are very close to
the numerical eigenvalues, indicating that the ansatz is a good guess
for the exact solution. There is however some small difference, as is
shown in  
Fig.~\ref{fig3} by comparing the above 
 ansatz with the numerically determined 
eigenfunctions in the case  $N=4$.

Finally, it is interesting to estimate the behaviour of the two-body
density matrix in
the thermodynamic limit, defined as the limit $N\to\infty,\ \omega\to
0$ keeping the central density
$n_{0}=N\phi_{0}^{2}(0)=N\sqrt{m\omega/\pi\hbar}$ constant. In that
case  Eq.~(\ref{eq:rho2}) takes the same form as in the homogeneous system \cite{GirMin05} and hence it follows that  
$\rho_{2}(x_{1},x_{2};x_{1}',x_{2}')=\lambda_{1}u_{1}(x_{1},x_{2})u_{1}(x_{1}',x_{2}')$ (apart from
terms decaying exponentially with pair separation),  with
$\lambda_{1}=N/2$ and $u_{1}(x_{1},x_{2})\propto 
\text{sgn}(x_{1}-x_{2})e^{-2n_{0}|x_{1}-x_{2}|}$. Hence, the estimate
in the thermodynamic limit agrees to order $1/N$ with the 
numerical solution.


%

\section{Summary and concluding remarks}
In summary, by estimating the largest eigenvalue of the two-body
reduced density matrix  we have investigated the possibility of
off-diagonal long-range order for a fermionic Tonks-Girardeau gas
subjected to a longitudinal harmonic  confinement. The result of
variational approach and numerical diagonalization together with an
estimate in the thermodynamic limit is that  the
largest eigenvalue $\lambda_1$ is of order  $N/2$, where $N$ is the particle number, and hence 
we find  a partial ODLRO. The  value  $\lambda_1\simeq N/2$ might also be
interpreted as the fraction of pairs which are Bose-Einstein condensed
in the one-dimensional equivalent of the BCS to BEC crossover for
$p$-wave fermions.
The emerging picture is the one
of  a partially paired
quantum state, where quantum fluctuations play a major role in
depleting the ``BEC'' of pairs. 
The consequences of pairing remain to be explored. 

\begin{acknowledgments}
This work was initiated at the Aspen Center for Physics during the
summer 2005 workshop ``Ultracold Trapped Atomic Gases''. We are grateful to the
Center staff and to the workshop organizers, Gordon Baym, Randy Hulet,
Erich Mueller, and Fei Zhou, for the opportunity to participate.
This paper has benefited from our conversations with other workshop
participants, particularly Stefano Giorgini, Robert Seiringer, Fei Zhou,
and Eugene Zaremba. The Aspen Center for Physics is supported in part by the
U.S. National Science Foundation, research of M.D.G. at the University of
Arizona by U.S. Office of Naval Research grant N00014-03-1-0427 through a 
subcontract from the University of Southern California, and that of
A.M.by the Centre Nationale de la Recherche Scientifique and by the Minist\`ere de la Recherche (grant ACI Nanoscience 201).
\end{acknowledgments}
\end{document}